\documentclass[letterpaper, 10 pt, conference]{ieeeconf}  % Comment this line out if you need a4paper

\IEEEoverridecommandlockouts                              % This command is only needed if 
\usepackage{amsthm}
\usepackage{graphics} % for pdf, bitmapped graphics files
\usepackage{epsfig} % for postscript graphics files
\usepackage{times} % assumes new font selection scheme installed
\usepackage{amsmath} % assumes amsmath package installed
\usepackage{amssymb}  % assumes amsmath package installed

\usepackage{biblatex}
\usepackage{bm}
\usepackage{xcolor}
\usepackage{graphicx}
\usepackage{algorithm}
\usepackage[noend]{algpseudocode}
\usepackage{booktabs}
\usepackage{siunitx}% An awesome package for typesetting and manipulation numbers and units.
\usepackage{hyperref}% Used for insert the link

\usepackage{caption}% Better control over caption
\usepackage{lipsum}% Example text
\usepackage{multirow}
\usepackage{diagbox} 
\usepackage{subcaption}

\newcommand{\R}{\mathbb{R}}%commands for easy math notations
\newcommand{\io}{\iota_{\varepsilon}}
\newcommand{\casr}{\mathcal{R}^{C,i,j}_{\varepsilon}}

\newcommand{\sigi}{\sigma_i}
\newcommand{\sigj}{\sigma_j}

\newtheorem{theorem}{\bf Theorem}
\newtheorem{lemma}{\bf Lemma}

\title{\LARGE \bf
Risk of Cascading Failures in Time-Delayed Vehicle Platooning$^{*}$
}

\author{Guangyi Liu, Christoforos Somarakis, and Nader Motee% <-this % stops a space

\thanks{
$^*$ This work was supported in parts by the AFOSR FA9550- 19-1-0004 and ONR N00014-19-1-2478.\endgraf
G. Liu and N. Motee are with the Department of Mechanical Engineering and Mechanics, Lehigh University, Bethlehem, PA, 18015, USA. {\tt\small \{gliu,motee\}@lehigh.edu}.
C. Somarakis is with the System Science Lab, Palo Alto Research Center, Palo Alto, CA, 94304. {\tt\small somarakis@parc.com.}
}

}

\begin{document}

\maketitle

% Replace this two for submission
\thispagestyle{plain}
\pagestyle{plain}

\begin{abstract} 
We develop a systemic risk framework to explore cascading failures in networked control systems. A time-delayed version of the vehicle platooning problem is used as a benchmark to study the interplay among network connectivity, system dynamics, physical limitations, and uncertainty onto the possibility of cascading failure phenomena. The measure of value-at-risk is employed to investigate the domino effect of failures among pairs of vehicles within the platoon. The systemic risk framework is suitably extended to quantify the robustness of cascading failures via a novel manipulation of bi-variate distribution. We establish closed-form risk formulas that explain the effect of network parameters (e.g., Laplacian eigen-spectrum, time delay), noise statistics, and systemic event sets onto the cascading failures. 
% We exploit the structure of a class of standard graphs (e.g., path, cyclic, regular graphs) and show how the risk of the second collision depends on the location of the first collision in the platoon. 
Our findings can be applied to the design of robust platoons to lower the cascading risk. We support our theoretical results with extensive simulations.
\end{abstract}

%%%%%%%%%%%%%%%%%%%%%%%%%%%%%%%%%%%%%%%%%%%%%%%%%%%%%%%%%%%%%%%%%%%%%%%%%%%%%%%%%%%%%

\section{Introduction}
The performance of networked control systems often suffers from communication or other physical limitations as well as external disturbances. This phenomenon is expected in the examples of autonomous vehicle platooning, synchronous power networks with integrated renewable sources, water supply networks, transportation networks, and the inter-dependent financial systems \cite{acemoglu2015systemic, bamieh2012coherence, bertsimas1998air, dorfler2012synchronization, kessler1989analysis}. In these applications, the external disturbance and communication limitations usually drive the networked system from the desired state of operation into failures.

Performance studies in networked control systems involve aggregate system response to an additive source of statistical uncertainty, followed by algorithmic methods for robust network synthesis. In essence, these design methods seek to minimize an aggregate input/output energy norm by appropriate calibration of feedback gain magnitude that in the context of multi-agent systems correspond to connectivity edges and weights. The $\mathcal H_2$ norm is a popular example of a systemic norm to minimize because it connects with elegant mathematics the aggregate state deviation and network topology characteristics in recent works \cite{bamieh2012coherence, siami2016fundamental, siami2017growing, ghaedsharaf2016interplay, grunberg2017determining, jovanovic2008peaking}. In other words, existing design methods target delivering robust networks where all nodes in the graph withstand perturbations that can deviate them from their nominal mode of operation.

The present paper explores an alternative path. Our working hypothesis is that faults and undesirable dynamics are inevitable in any real-world systems. For interconnected systems in particular, the issue is that the slightest local misbehavior can propagate through systems' interaction mechanisms resulting in global failure. This navigates our interests to the study of cascading failures  \cite{7438924, zhang2018cascading, zhang2019robustness}. From our perspective, it is meaningful to investigate system design tools in networked control systems that characterize networks' ability to isolate the propagation of a failure that has already occurred within the network component. We will develop a theoretical framework based on the systemic risk analysis to evaluate the possibility of cascading (domino) failures in the network. Our objective is to highlight how the impact of a systemic failure in some areas of the network can trigger other failures throughout the network. Providing the quantification of such impact will give valuable insights into the system design and highlight the potential of a networked system to be suffered from the cascading systemic failures.

Our case study is a platoon of identical autonomous vehicles that communicates over a time-invariant network with lagged transmission and processing of information in agents' sensors and actuators. The network is perturbed by statistical noise applied on each node independently. This noise models the effect of external perturbations and turns the system into a stochastic dynamical network. Our focus is on the study of events of collision in pairs of successive vehicles, given that another pair of vehicles in the platooning has already collided. 

This is, to the best of our knowledge, the first formal approach to quantify the robustness of cascading failure events in large-scale dynamical networks with deficiencies. Our preliminary results indicate radical differences in the system components that contribute to cascading failures from mainstream results. This motivates us to turn our research efforts to design methods for networks that aim to isolate the existing internal failures rather than simply withstand them as external disturbances.

{\it Our Contributions: }
Building upon our recent works on the first- and second-order linear consensus networks \cite{Somarakis2016g,Somarakis2017a, Somarakis2019g,Somarakis2020b, somarakis2018risk}, this work extends our results on the risk measure of single systemic events into cascading systemic events. 
% Cascading events, which are usually expressed as conditional events, are different from joint events. 
We examine the value-at-risk of cascading failures in the steady-state distribution of a time-delayed vehicle platooning. In particular, we obtain explicit formulas for the risk of collision of a pair of vehicles given that another pair of vehicles has already collided. We explain the contribution of nodes' and interacting nodes' uncertainties on the marginal variance and correlation between the corresponding pairs of vehicles, respectively.
Furthermore, we explore the problem of risk-based design methods by aiming to minimize the risk of cascading failures for certain graphs.
% We highlight some time-delay induced fundamental (lower) limits on the risk of cascading events for those graphs, and propose a potential generalization of those limits to arbitrary communication graphs.
Finally, we use extensive simulations to validate our theoretical findings. All proofs of theorems and lemmas can be found in the appendix.

%%%%%%%%%%%%%%%%%%%%%%%%%%%%%%%%%%%%%%%%%%%%%%%%%%%%%%%%%%%%%%%%%%%%%%%%%%%%%%%%%%%%%
\section{Preliminaries} We follow the standard notation for the set of real valued scalars, vectors and matrices. The set of all non-negative real numbers is denoted by $\mathbb{R}_{+}$, and the set of all positive real numbers is presented by $\mathbb{R}_{++}$. We denote the vector of all ones by $\bm{1}_n = [1, \dots, 1]^T \in \mathbb{R}^{n}$. The set of standard Euclidean basis of $\mathbb{R}^{n}$ is represented by ${\bm{e}_1, \dots, \bm{e}_n}$, and we denote $\tilde{\bm{e}}_i = \bm{e}_{i+1} - \bm{e}_{i}$ for all $i \in \{1, \dots, n-1\}$. 

{\it Algebraic Graph Theory:} We define an undirected graph by $\mathcal{G} = (\mathcal{V}, \mathcal{E}, \omega)$, in which $\mathcal{V}$ denotes the set of all nodes, $\mathcal{E}$ denotes the set of all edges (links), and $\omega: \mathcal{V} \times \mathcal{V} \rightarrow \mathbb{R}_{+}$ represents the weight function that assign all edges $(i,j)$ with weight values $\omega_{i,j} > 0$. In this paper, we assume every graph is connected.
% \begin{assumption}  \label{asp:connected}
%     Every graph in this paper is connected.
% \end{assumption}
The Laplacian matrix of a graph $\mathcal{G}$ is $L = [l_{ij}] \in \R^{n \times n}$ with elements
\[
    l_{ij}=\begin{cases}
        \; -\omega_{i,j}  &\text{if } \; i \neq j \\
        \; \sum_{j = 1}^{n} \omega_{i,j}  &\text{if } \; i = j 
    \end{cases}.
\]

It is known that the Laplacian matrix of a graph is symmetric and positive semi-definite \cite{van2010graph}. The assumption of a connected graph implies the smallest Laplacian eigenvalue is zero with algebraic multiplicity one. The spectrum of $L$ can be ordered as 
$
    0 = \lambda_1 < \lambda_2 \leq \dots \leq \lambda_n,
$
and their corresponding eigenvector is denoted by $\bm{q}_{k}$. Letting $Q = [\bm{q}_{1} | \dots | \bm{q}_{n}]$, we have $L = Q \Lambda Q^T$ and $\Lambda = \text{diag}[0, \lambda_2, \dots, \lambda_n]$. The eigenvectors are normalized such that $Q$ is an orthogonal matrix, i.e., $Q^T Q = Q Q^T = I_{n}$ with $\bm{q}_1 = \frac{1}{\sqrt{n}} \bm{1}_n$. 
% The connectivity of $\mathcal{G}$ is measured with the total effective resistance, which is characterized in \cite{klein1993resistance} as
% \begin{equation}
%     \Xi_{\mathcal{G}} = n \sum^{n}_{i=2} \lambda_{i}^{-1}.
% \end{equation}
% The smaller value of $\Xi_{\mathcal{G}}$ indicates a stronger connectivity of $\mathcal{G}$.

{\it Probability Theory:} We denote the set of $\mathbb{R}^{q}$ valued random variable as $\mathcal{L}^{2}(\mathbb{R}^{q})$ with the probability space $(\Omega, \mathcal{F}, \mathbb{P})$ and finite second moments. 
A normal random variable $\bm{y} \in \mathbb{R}^{q}$ with mean $\bm{\mu} \in \mathbb{R}^{q}$ and covariance matrix $\Sigma \in \mathbb{R}^{q \times q}$ is denoted by $\bm{y} \sim \mathcal{N}(\bm{\mu}, \Sigma)$. When $q = 2$, we say that $\bm{y}$ follows a bi-variate normal distribution \cite{tong2012multivariate}, the properties of which will be analyzed in \S \ref{bivariate}. The error function $\text{erf} (x): \R \rightarrow (-1,1)$ is 
$
\text{erf} (x) = \frac{2}{\sqrt{\pi}} \int_{0}^{x} e^{-t^2} \text{d} t,
$
which is invertible on its range.

%%%%%%%%%%%%%%%%%%%%%%%%%%%%%%%%%%%%%%%%%%%%%%%%%%%%%%%%%%%%%%%%%%%%%%%%%%%%%%%%%%%%%
\section{Problem Statement}\label{problemstatement}
\begin{figure}[t]
    \centering
	\includegraphics[width=\linewidth]{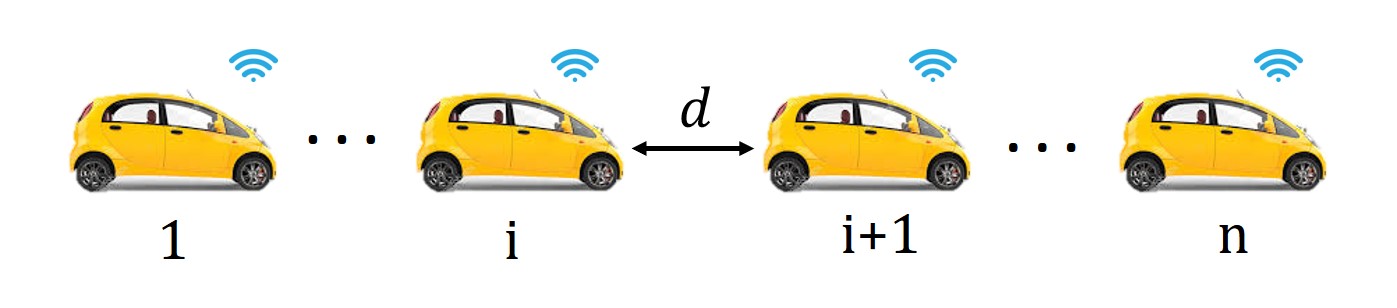}
	\caption{Schematics of platoon ensemble of autonomous vehicles. Speed and distance control is adjusted automatically with feedback laws using information communicated over a virtual network. }
	\label{fig:platoon}
\end{figure}

Let us consider a platoon of vehicles along the horizontal axis, that consists of $n$ vehicles (see Fig. \ref{fig:platoon}). The vehicles are labeled in a descending order such that the $n^{th}$ vehicle is considered as the leader of the platoon. The state of the $i^{th}$ vehicle is denoted by $[x^{(i)}_{t}, \text{v}^{(i)}_{t}]^T$, in which $x^{(i)}_{t}, \text{v}^{(i)}_{t} \in \mathbb{R}$  represent the position and velocity of the $i^{th}$ vehicle at time $t$, respectively. The dynamics of the $i^{th}$ vehicle is governed by stochastic differential equations \cite{Somarakis2020b}
\begin{equation} \label{equ:dyn_vehicle}
    \begin{aligned}
        d x^{(i)}_{t} &= \text{v}^{(i)}_{t} dt\\
        d \text{v}^{(i)}_{t} &= u^{(i)}_{t} dt + g\, d \xi^{(i)}_{t}
    \end{aligned}
\end{equation}
where $u^{(i)}_{t}$ is the control input. We model the uncertainty in the dynamics of each vehicle with $g\, d \xi^{(i)}_{t},\; i \in \mathcal{V}$, which represent the white noise generators\footnote{The stochastic process $\xi^{(i)}_{t} \in \mathcal{L}^2(\R)$ denotes the real-valued Brownian motion.}. It is assumed that the white noise is additive and it is independent of other vehicles' noises. The magnitude of the noise is measured by a diffusion parameter $g \in \mathbb{R}_{++}$, and it is assumed to be identical among all vehicles. The objectives of the control of the platoon are to guarantee the following two global behaviors: (i) the pair-wise differences between position variables $x^{(i)}_{t}$ of every two consecutive vehicles converge to 0; and (ii) all vehicles attain the same velocity in the steady-state. Assuming that for any information to be communicated and processed, there is non-negligible time-delay, modeled as a lumped parameter $\tau \in \mathbb{R}_{++}$. It is known from \cite{yu2010some} that the following feedback control law can achieve objectives (i) and (ii)
\begin{equation} \label{equ:feedback}
    \begin{aligned}
        u^{(i)}_{t} = \sum_{j=1}^{n} &\omega_{i,j} \Big( \text{v}^{(j)}_{t - \tau} - \text{v}^{(i)}_{t - \tau}\Big)\\
        &+ \beta \sum_{j=1}^{n} \omega_{i,j} \Big( x^{(j)}_{t - \tau} - x^{(i)}_{t - \tau} - (j-i) d \Big).
    \end{aligned}
\end{equation}
The parameter $\beta \in \R_{++}$ balances the effect of the relative positions and the velocities. We define the vector of positions, velocities, and noise inputs as $\bm{x}_t = [x_{t}^{(1)}, \dots, x_{t}^{(n)}]^T$, $\textbf{v}_t = [\text{v}_{t}^{(1)}, \dots, \text{v}_{t}^{(n)}]^T$ and $\bm{\xi}_t = [\xi_{t}^{(1)}, \dots, \xi_{t}^{(n)}]^T$. We denote the target distance vector as $\bm{y} = [d,2d,\dots,nd]^T$. Applying \eqref{equ:feedback} to \eqref{equ:dyn_vehicle}, we represent the closed-loop dynamics as an initial value problem
\begin{equation}\label{equ:dyn}
    \begin{aligned}
        d\bm{x}_t &= \textbf{v}_t dt,\\
        d\textbf{v}_t &= -L \textbf{v}_{t-\tau} dt - \beta L (\bm{x}_{t-\tau} - \bm{y}) dt+ g d \bm{\xi}_t,
    \end{aligned}
\end{equation}
for all $t\geq 0$ and given deterministic values of $\bm{x}_t$ and $\textbf{v}_t$ for $t \in [-\tau, 0]$. This poses as stochastic functional differential equations \cite{mohammed1984stochastic, Evans2013}, the standard result guarantees that $\{(\bm{x}_t, \textbf{v}_t)\}_{t\geq -\tau}$ is a well-posed stochastic process.

The {\it problem} is to quantify the risk of cascading failures as a function of the underlying graph Laplacian, time-delay, and statistics of noise under the condition when a similar systemic failure has already occurred in another location in the system. To this end, we will develop a systemic risk framework to study cascading failures of events based on the steady-state statistics of the closed-loop system  \eqref{equ:dyn_vehicle} and \eqref{equ:dyn}. The rest of the paper is organized as follows. In \S \ref{prelims} we collect a few key preliminary results on the steady-state behavior and statistics of the closed-loop system, as well as some useful observations on bi-variate normal distributions. These results help us shape the risk formula of cascading failures in the \S \ref{risk} to follow, which constitutes the main contribution of this work. \S \ref{limits} is devoted to designing challenges for risk minimization with general remarks. Special graph topologies are examined in \S \ref{sec:case-study} where we show by simulation the complexity of cascading failure phenomena.

%%%%%%%%%%%%%%%%%%%%%%%%%%%%%%%%%%%%%%%%%%%%%%%%%%%%%%%%%%%%%%%%%%%%%%%%%%%%%%%%%%%%%
\section{Preliminary Results}\label{prelims}
In this section, we present the conditions of deterministic stable platoon dynamics, and the steady-state statistics of the distance between vehicles with external disturbances. Such statistics constitute the latter results of a useful observation on the bi-variate normal distribution and the value-at-risk measure. 

\subsection{Internal Stability}
We denote the convergence of the platoon as the velocities of all vehicles obtain the same value and the distances among all consecutive vehicles are $d \in \R_{++}$, which is equivalent to 
\begin{equation*}
    \lim_{t \rightarrow \infty} |\text{v}_{t}^{(i)} - \text{v}_{t}^{(j)}| = 0 ~\text{and}~ \lim_{t \rightarrow \infty} |x_{t}^{(i)} - x_{t}^{(j)} - (i-j) d| = 0
\end{equation*}
for all $i,j \in \mathcal{V}$. A recent work \cite{yu2010some} has shown that the deterministic platoon will converge if and only if 
\begin{align}   \label{equ:converge}
    (\lambda_i \tau, \beta \tau ) \in S,
\end{align}
in which we consider the set
\begin{align*}
    S = \bigg\{(s_1,s_2) \in \R^2 \big| s_1 \in (0, \frac{\pi}{2}), s_2 \in (0, \frac{a}{\tan(a)}, \\\text{for } a \in (0,\frac{\pi}{2})) \text{ the solutions of } a\sin(a) = s_1 \bigg\}.
\end{align*} 
In this work, we only consider the cases when the deterministic platoon dynamic is convergent, i.e., $(\lambda_i \tau, \beta \tau ) \in S$.

\subsection{Steady-State Statistics of Distances}\label{ssstatistics}
Using the decomposition of $L = Q \Lambda Q^T$, \eqref{equ:dyn} can be transformed into another coordinate by the following transformation
$$
    \bm{z}_t = Q^T (\bm{x}_t -\bm{y}) \text{, and } \bm{\upsilon}_t = Q^T \textbf{v}_t.
$$
The platoon system in the new coordinate is shown by
\begin{equation}\label{equ:dyn_z}
    \begin{aligned}
        \text{d}\bm{z}_t &= \bm{\upsilon}_t \;\text{d}t,\\
        \text{d}\bm{\upsilon}_t &= -\Lambda \bm{\upsilon}_{t-\tau} \;\text{d}t \; - \; \beta \Lambda \bm{z}_{t-\tau} \;\text{d}t \;+\; g Q^T \; \text{d} \bm{\xi}_t.
    \end{aligned}
\end{equation}
Once the stability condition of \eqref{equ:dyn_z} is satisfied, the solution of decoupled subsystems can be represented as
\begin{equation}
    \begin{aligned}
        \begin{bmatrix}
        z_t^{(i)} \\
        v_t^{(i)} 
        \end{bmatrix} = 
       \Gamma(z_{[-\tau,0]};v_{[-\tau,0]}) + g\int^{t}_{0} \Phi_i(t-s) B_i d\bm{\xi}_s.
    \end{aligned}
\end{equation}
for $\Gamma(\cdot \,; \cdot )$ a generalized function describing transient dynamics that vanish exponentially fast \cite{Somarakis2020b}, and $B_i = [\bm{0}_{1 \times n}, \bm{q}_i^T]^T$. It is known from \cite{Somarakis2020b} that the statistics of the steady-state $\bar{\bm{z}} \in \R^{n}$ follows a multi-variate normal distribution with mean $0 \cdot \bm{1}_n$. Its covariance matrix is shown by
$$
    \Sigma_z = \text{diag} \Big\{ \sigma_{z_1}^2, ..., \sigma_{z_n}^2\Big\},
$$
% in which $\sigma_{z_i}^2 = g^2 \int_{0}^{t} \tilde{\Phi}_{ii}(t-s) ds$, and 
% \begin{align*}
%     \tilde{\Phi}(t-s) = \text{diag} \bigg\{ 
%     \begin{bmatrix}
%     0 &1
%     \end{bmatrix} \Phi_i(t-s) 
%     \begin{bmatrix}
%     0 &0\\
%     0 &1
%     \end{bmatrix}\Phi_i(t-s)^T
%     \begin{bmatrix}
%     0\\
%     1
%     \end{bmatrix}
%     \bigg\}.
% \end{align*}
% Take the limit as $t \rightarrow \infty$, one can have 
in which $\sigma_{z_i}^2 = \frac{g^2 \tau^3}{2\pi}$ $ f(\lambda_i\tau, \beta \tau)$ with 
\begin{equation} \label{equ:f}
    f(s_1, s_2) = \int_{\R} \frac{\text{d}\,r}{(s_1s_2 - r^2 \cos(r))^2 + r^2 (s_1-r \sin(r))^2}.
\end{equation}

Given that $\bm{x}_{t} = Q \bm{z}_{t} + \bm{y}$, we define the steady-state distance vector of the platoon as $\bm{\bar{d}} \in \mathbb{R}^{n-1}$ such that
\begin{align}\label{equ:dist}
    \bm{\bar{d}} = D Q \bm{\bar{z}} + d \bm{1}_n,
\end{align}
in which $D = \big[\tilde{\bm{e}}_1^T \big| \dots \big|\tilde{\bm{e}}_{n-1}^T \big]^T$ and $Q = [ \bm{q}_1 | \dots | \bm{q}_n]$. The $i^{th}$ element in $\bm{\bar{d}}$ denotes the distance between the $i^{th}$ and $(i+1)^{th}$ vehicle in the platoon for all $i \in \{1,\dots, n-1\}$.
 
\begin{theorem}     \label{thm:d_steady}
    Suppose the stability condition \eqref{equ:converge} holds, the steady-state inter-vehicle distance vector $\bar{\bm{d}}$ follows a multi-variate normal distribution in $\R^{n-1}$ such that
    \begin{align*}
        \bm{\bar{d}} \sim \mathcal{N}(d \bm{1}_n, \Sigma),
    \end{align*}
    with the elements of $\Sigma = [\sigma_{ij}]$ is shown by
    \begin{equation} \label{equ:sigma_d}
    \begin{aligned}
        \sigma_{ij} = 
        g^2 \frac{\tau^3}{2\pi} \sum_{k=1}^{n} \big(\tilde{\bm{e}}_{i}^T \bm{q}_k \big) \big(\tilde{\bm{e}}_{j}^T \bm{q}_k \big) f(\lambda_k \tau, \beta \tau),
    \end{aligned}
    \end{equation}
    for all $i,j = \{1,\dots, n-1\}$, and $f(\lambda_k \tau, \beta \tau)$ as in \eqref{equ:f}. We also denote the diagonal elements $\sigma_{ii}$ as $\sigi^2$. 
\end{theorem}

\subsection{Bi-variate Normal Distribution}\label{bivariate}
In order to evaluate the impact of a systemic failure on other vehicles in the system, it is immediate to consider the statistics of the other vehicles' states with the conditional probability by assuming one systemic failure has occurred. Hence, we provide a closed-form representation of the statistics of the steady-state distance at the $j^{th}$ pair $\bar{d}_j$ when the systemic failure has occurred at the $i^{th}$ pair of vehicles. 

\begin{lemma}   \label{lemma:conditional_prob}
    For any pair of two steady-state distance vectors $[\bar{d}_i, \bar{d}_{j}]^T$ with $i,j \in \{1,\dots, n-1\}$ and $i \neq j$, they follow a bi-variate normal distribution. Given that $\bar{d}_{i} = d_c$, one can obtain the conditional distribution of $\bar{d}_{j}$ as a normal distribution $\mathcal{N}(\tilde{\mu},\tilde{\sigma}^2)$ with the mean 
    $$\tilde{\mu} = d + \rho_{ij} \frac{\sigj}{\sigi}(d_c - d),$$ 
    and the variance 
    $$\tilde{\sigma}^2 = \sigj^2 (1 - \rho_{ij}^2),$$ 
    in which $\rho_{ij} = \sigma_{ij} / \sigma_{i} \sigma_{j}$ and $|\rho_{ij}| < 1$. We refer $\sigma_{ij}, \sigma_{i}, \text{ and }\sigma_{j}$ to \eqref{equ:sigma_d}.
\end{lemma}

We represent the systemic failures with the value of $d_c$, e.g., a collision event implies $d_c = 0$. The above lemma provides the statistics of the distance vector knowing the existence of the systemic failure, and this result can be applied to any two pairs of vehicles. 
Observing \eqref{equ:sigma_d}, one can find out that the cross-correlation $\rho_{ij}$ is independent of the magnitude of external noise. This implies that the probabilistic inter-relation between two pairs of vehicles originated from the time-delay and the communication graph but not the stochastic perturbation.

\subsection{Value-at-Risk Measures}

The value-at-risk measure \cite{Follmer2016,rockafellar2000optimization} has been shown to be a valid measure for the systemic risk in the vehicle platooning (see \cite{Somarakis2016g,Somarakis2020b}). These risk measures quantify the chance of a random variable (e.g., $\bar{d}_j$) is steered into an undesirable ranges of values, denoted by $U$. We show the undesirable values, which is also referred to the collection of systemic sets, as $\{U_{\delta}\}$ parametrized by $\delta \in [0, \infty]$. The systemic set is defined in the manner that it enjoys the following properties
\begin{itemize}
    \item $U_{\delta_1} \subset U_{\delta_2}$ when $\delta_1 > \delta_2$.
    \item $\lim_{n \rightarrow \infty} U_{\delta_n} = \bigcap_{n=1}^{\infty} U_{\delta_n} = U$ for any sequence $\{\delta_n\}^{\infty}_{n=1}$ with $\lim_{n \rightarrow \infty} \delta_n = \infty$.
\end{itemize}
For a real-valued random variable $y$ with probability space $(\Omega, \mathcal{F}, \mathbb{P})$, we define the systemic event as $\{y \in U \}$, and the scalar value-at-risk measure $\mathcal{R}_{\varepsilon} : \mathcal{F} \rightarrow \R_{+}$ is given by
$$
    \mathcal{R}_{\varepsilon} = \inf \big\{ \delta>0 \big| \mathbb{P} \{y \in U_{\delta} \} < \varepsilon \big\}.
$$
The parameter $\varepsilon \in (0,1)$ denotes the level of confidence in the systemic events (e.g., collision). The smaller this value, the higher the confidence of the random variable $y$ stays away from the systemic set $U$. The value-at-risk measure, $\mathcal{R}_{\varepsilon}$, represents the intuitive notion of "risk." The higher its value, the higher chance the system will be steered into the undesirable ranges of values.

%%%%%%%%%%%%%%%%%%%%%%%%%%%%%%%%%%%%%%%%%%%%%%%%%%%%%%%%%%%%%%%%%%%%%%%%%%%%%%%%%%%%%
\section{Risk of Cascading Inter-Vehicle Collision}\label{risk}
For the exposition of the next results, we assume the platooning ensemble travels along a straight line, as illustrated in Figure \ref{fig:platoon}. Naturally, there is a sequential enumeration ${1,\dots,n-1}$ for pairs of consecutive vehicles. Following notation of Lemma \ref{lemma:conditional_prob}, let $i^{th}$ and $j^{th}$ be two pairs of vehicles in the ensemble, where $i^{th}$ pair means vehicles $i$ and $i+1$ and so forth. The main scenario assumes a collision to have occurred among the $i^{th}$ pair of vehicles, and we study the risk of collision at the $j^{th}$ pair in view of the latter event. 

We define the event of under the \textit{risk of collision} for the  steady-state distance $\bar{d}_{j}$ as 
\begin{align*}
    \Big\{ \bar{d}_{j} \in C_{\delta} \Big\} \text{ with } C_{\delta}=\Big( -\infty, \frac{d}{\delta+c} \Big),
\end{align*}
for $\delta \in [0,\infty]$ and $c \geq 1$. The associated risk measure of cascade collision is defined with the conditional probability distribution of $\bar{d}_{j}$ assuming the $i^{th}$ pair has collided, i.e., $\bar{d}_{i} = 0$. The value-at-risk measure for cascading collision is defined as
\begin{align}\label{equ:var}
    \mathcal{R}^{C,i,j}_{\varepsilon} = \inf \big\{ \delta > 0 \;\big|\; \mathbb{P} \{ \bar{d}_{j} \in C_{\delta} \;\big|\; \bar{d}_i = 0\} < \varepsilon \big\}
\end{align}
with the confidence level $\varepsilon \in (0,1)$. The larger the value of risk, the higher the probability of the cascading collision is going to occur for the system. For the exposition of the next result we introduce the notation:
$$\kappa_{\delta}^{(i,j)}: = \frac{d}{\sqrt{2(1-\rho_{ij}^2)}\, \sigma_j} \bigg( \frac{1}{\delta + c}+ \rho_{ij} \frac{\sigma_j}{\sigma_i} -1 \bigg)$$ and $$\gamma(i,j,\varepsilon) = \iota_{\varepsilon} \sigma_{j}\sigma_{i} \sqrt{2(1-\rho_{ij}^2)}+d\,\sigma_i-d\,\rho_{ij}\,\sigma_j$$ 
with $\sigma_i, \sigma_{j}, \rho_{ij}$, for the $i^{th},j^{th}$ pair of vehicles in a platoon, their associated standard deviations and correlation coefficient. It is noted that $\kappa$ is a function parametrized by $\delta\in [0,\infty]$ and taking values in $[\kappa_0^{(i,j)} , \kappa_{\infty}^{(i,j)}]$ defined for $|\rho_{ij}|<1$.

\begin{theorem}\label{thm:cascade_risk_collision}
    Suppose that the conditions of stability hold and the $i^{th}$ pair has collided. For $i, j \in \{1, \dots, n-1\}$ and $i \neq j$, the risk of cascading inter-vehicle collision at the $j^{th}$ pair is 
    \[
        \mathcal{R}^{C,i,j}_{\varepsilon}=\begin{cases}
            0, &\text{if} ~ \kappa_{0}^{(i,j)} \leq \iota_{\varepsilon}\\
            \dfrac{d\,\sigma_i}{\gamma(i,j,\varepsilon)} - c, &\text{if} ~ \iota_{\varepsilon} \in \big(\kappa_{\infty}^{(i,j)},\kappa_{0}^{(i,j)} \big) \\
            \infty, &\text{if} ~ \kappa_{\infty}^{(i,j)} \geq  \iota_{\varepsilon}
            \end{cases}
    \]
    where $\iota_{\varepsilon} = \text{\normalfont erf}^{-1} (2 \varepsilon - 1)$.
\end{theorem}

The above theorem addresses the value-at-risk for the cascading collision with the branches defined by $\kappa_{0}^{(i,j)},\kappa_{\infty}^{(i,j)}$, and $\io$. The two extreme cases are self-explanatory: If $\kappa_{0}^{(i,j)} \leq \iota_{\varepsilon}$, the probability of the $j^{th}$ pair entering the systemic set $C_{\delta}$ is strictly below $\varepsilon$ with an arbitrary selection of $\delta$. On the other extreme, the $j^{th}$ pair will enter $C_{\delta}$ with the probability at least $\varepsilon$ if $\kappa_{\infty}^{(i,j)} \geq  \iota_{\varepsilon}$. 

When there is no cross-correlation between two pairs of vehicles, i.e., $\rho_{ij}=0$, the second branch of $\casr$ boils down to the risk of a single collision at the $j^{th}$ pair of vehicles, 
\begin{equation} \label{equ:single}
    \mathcal{R}^{C,j}_{\varepsilon} = \frac{d}{\io \sqrt{2} \sigj + d} -c, 
\end{equation}
when $\io \in (-\frac{d}{\sqrt{2} \sigj}, \frac{d}{\sqrt{2} \sigj} \frac{1-c}{c} )$. This is equivalent to the result presented in \cite{Somarakis2020b}. Given the $i^{th}$ pair of vehicles has collided, one can formulate the risk profile of the platoon by stacking up all risks of cascading collision among remaining pairs with a  systemic risk vector $\bm{\mathcal{R}}^{C,i}_{\varepsilon} \in \R^{n-2}$ as follows
\begin{equation*}
    \bm{\mathcal{R}}^{C,i}_{\varepsilon} = \big[\mathcal{R}^{C,i,1}_{\varepsilon}, \dots, \mathcal{R}^{C,i,i-1}_{\varepsilon}, \mathcal{R}^{C,i,i+1}_{\varepsilon}, \dots , \mathcal{R}^{C,i,n-1}_{\varepsilon} \big]^T.
\end{equation*}

\begin{figure*}[t] 
    \begin{subfigure}[t]{.28\linewidth}
        \centering
    	\includegraphics[width=\linewidth]{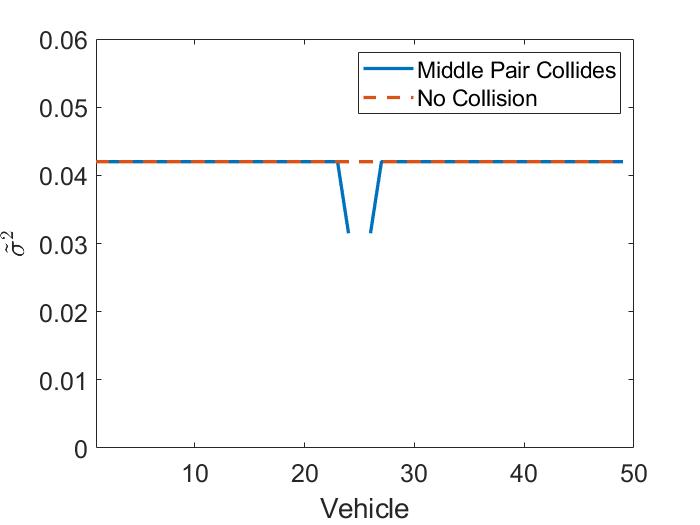}
    	\caption{The complete Graph.}
    \end{subfigure}
    \hfill
    \begin{subfigure}[t]{.28\linewidth}
        \centering
    	\includegraphics[width=\linewidth]{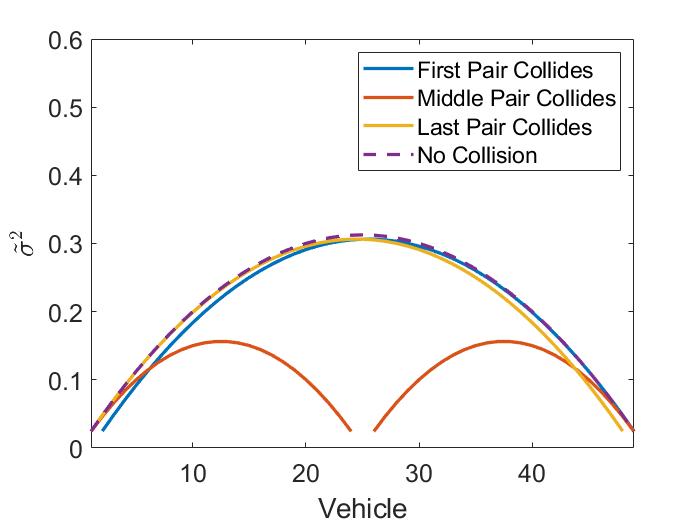}
    	\caption{The path graph.}
    \end{subfigure}
    \hfill
    \begin{subfigure}[t]{.28\linewidth}
        \centering
    	\includegraphics[width=\linewidth]{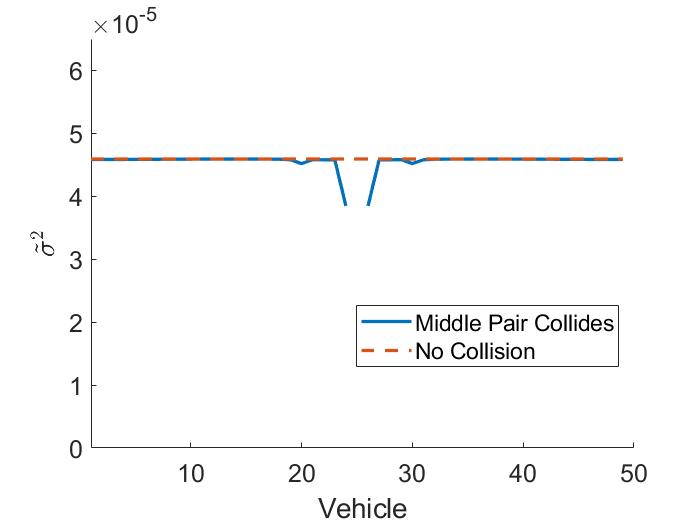}
    	\caption{The p-cycle graph with $p = 5$}
    \end{subfigure}
    \caption{The plot of $\tilde{\sigma}^2$ with and without the existence of collisions.}
    \label{fig:variance}
\end{figure*}

%%%%%%%%%%%%%%%%%%%%%%%%%%%%%%%%%%%%%%%%%%%%%%%%%%%%%%%%%%%%%%%%%%%%%%%%%%%%%%%%%%%%%
\section{Risk Design Objectives and Challenges}\label{limits}
The central problem in control synthesis is to deliver controls that optimize the plant towards certain objectives. In the present work, the objective function is the risk of cascading failures as described in Theorem \ref{thm:cascade_risk_collision}. The question is, how can one unveil a design strategy by addition, removal, or edge calibration (weight adjustment) steps that achieve a smaller risk value? Is there an optimal risk value? It turns out that the minimization of cascading risk is a very challenging problem. To see this, recall that in $\rho_{ij}=0$, the risk formula in Theorem \ref{thm:cascade_risk_collision} simplifies to the risk of single collision as in \eqref{equ:single}. It is obvious that the risk is then a monotonic function of $\sigma_{j}$ that in turn attains a time-delay induced fundamental limit \cite{Somarakis2020b}. However, the risk of cascading failure involves, in addition to $\sigma_j$, the uncertainty of the $i^{th}$ pair together with the correlation $\rho_{ij}=\frac{\sigma_{ij}}{\sigma_i\, \sigma_j} \in (-1,1)$, that characterizes the dependency of the corresponding pairs of vehicles.  

Let us focus on risk with acceptance level $\varepsilon \in (0,1/2)$. Elementary manipulation of risk formula shows that risk-constraint design involves the minimization of 
$$ \mathcal K(\sigma_i,\sigma_j,\sigma_{ij}) := \frac{|\iota_{\varepsilon}| \sqrt{2}}{d}\sqrt{\sigma_{j}^2-\frac{\sigma_{ij}^2}{\sigma_{i}^2}}+\frac{\sigma_{ij}}{\sigma_i^2}, $$ 
with respect to $\sigma_i,\sigma_j$, and $\sigma_{ij}$ in \eqref{equ:sigma_d}. The objective of the design is in fact to minimize $\mathcal{K}$ by adjusting the communication graph together with \eqref{equ:sigma_d}.

Furthermore, depending on either $\mathcal K \geq 1$, $1>\mathcal K>1-1/c$, or $1-1/c \geq \mathcal K$, the risk value is infinite, positive, or zero. While the universal lower bound of $\sigma_i$ restricts, as a consequence, the minimization of the single collision risk is not the case for $\mathcal K$. Heuristically speaking, if the network topology yields pair $j$ and $i$ to attain a correlation coefficient close enough to $-1$, then regardless of the value of $\sigma_j$, the risk of cascading collision can be zero. The interpretation is that $\rho_{ij} \approx -1$ implies a strong anti-correlation between the $j^{th}$ and the $i^{th}$ pair, and when relative distance of the $i^{th}$ pair decreases, the distance of the $j^{th}$ pair must increase, hindering the collision. With this last remark, we conclude that the risk of cascading failure exhibits a much more involved behavior that requires further study along the lines of understanding the role of $\sigma_{ij}$ as a function of graph coupling weights.

%%%%%%%%%%%%%%%%%%%%%%%%%%%%%%%%%%%%%%%%%%%%%%%%%%%%%%%%%%%%%%%%%%%%%%%%%%%%%%%%%%%%%
\section{Case Studies} \label{sec:case-study}
We discuss the case studies for platoons with dynamics governed by \eqref{equ:dyn_vehicle} and \eqref{equ:feedback} among the complete graphs, the path graphs, and the p-cycle graphs \cite{van2010graph}. The risk of cascading collision $\casr$ is computed under the assumption that the first, middle, or last pair have collided. We only show the collision at the middle pair for the complete and the $p-$cycle graph since the results are symmetric.

\subsection{Variance after Systemic Failure}
The impacts from collisions are evident through the conditional distribution of the steady-state distances $\bar{d}_j$ for all $j \in \{1, \dots, n-1\}$. We evaluate the conditional variance $\tilde{\sigj}^2$ as in Lemma \ref{lemma:conditional_prob} (see Fig \ref{fig:variance}). In all cases, the variance of conditional distribution $\tilde{\sigj}^2$ is upper bounded by the variance $\sigj^2$ when there is no systemic failure. That is, the distances of the remaining pairs tend to have a narrower spread when collision occurs. The result is self-explanatory since the overall uncertainty of the platoon decreases as we introduce the additional information, i.e., the fact that the collision occurs.

\subsection{Risk of Cascading Collision}
\begin{figure}[t]
    \begin{subfigure}[t]{.49\linewidth}
        \centering
    	\includegraphics[width=\linewidth]{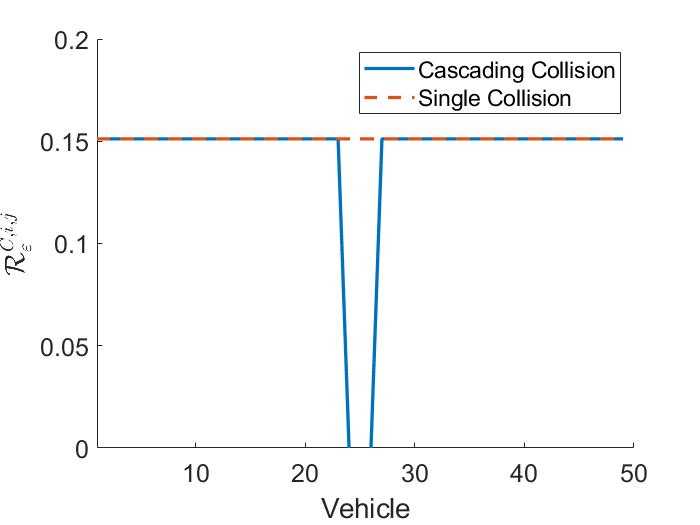}
    	\caption{The complete Graph.}
    	\label{fig:risk_collide_50}
    \end{subfigure}
    \hfill
    \begin{subfigure}[t]{.49\linewidth}
        \centering
    	\includegraphics[width=\linewidth]{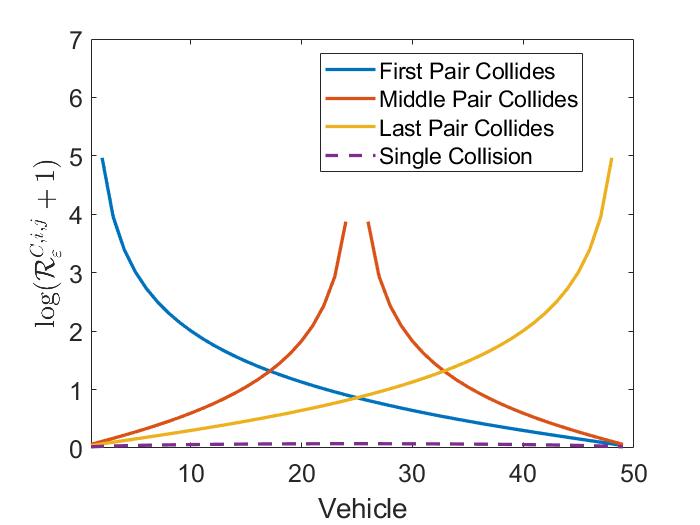}
    	\caption{The path graph.}
    	\label{fig:risk_collide_path_10}
    \end{subfigure}
    \medskip
    \begin{subfigure}[t]{.49\linewidth}
        \centering
    	\includegraphics[width=\linewidth]{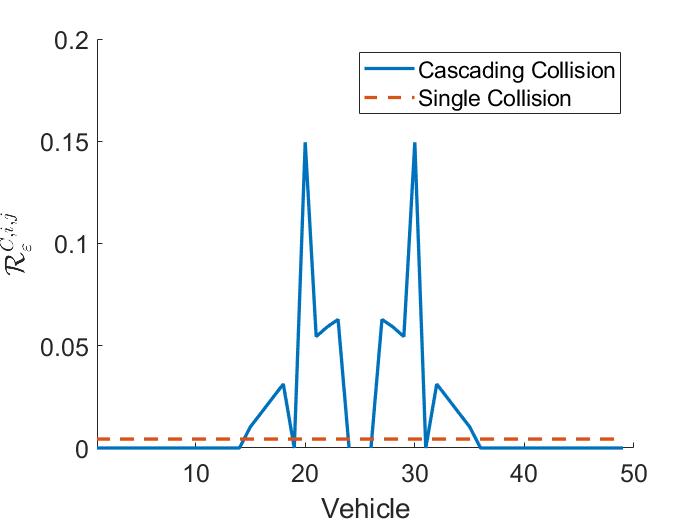}
    	\caption{The 5-cycle graph.}
    \end{subfigure}
    \hfill
    \begin{subfigure}[t]{.49\linewidth}
        \centering
    	\includegraphics[width=\linewidth]{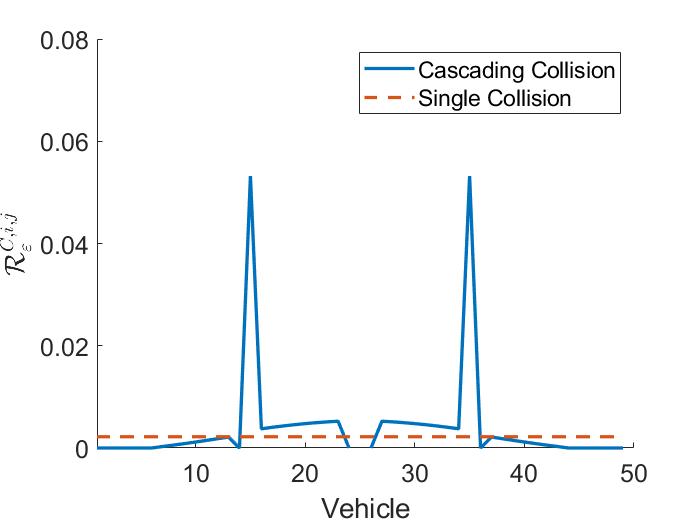}
    	\caption{The 10-cycle graph.}
    \end{subfigure}
    \caption{The plot of the risk of cascading collision $\casr$.}
    \label{fig:risk_collision}
\end{figure}

We evaluate the value-at-risk of cascading collision $\casr$ for all pairs of vehicles in the platoon (see Fig. \ref{fig:risk_collision}). We also show the value-at-risk of a single collision $\mathcal{R}^{C,j}_{\varepsilon}$ previously explored in \cite{Somarakis2020b} to serve as a benchmark. 

\subsubsection{Complete Graph}
We set $n = 50, g = 10, \tau = 0.02,\beta = 1, c = 1, d = 2, \text{ and } \varepsilon = 0.1$. It is shown that for the immediate neighbor of the collision, $\casr$ obtains a lower value than the risk of a single collision $\mathcal{R}^{C,j}_{\varepsilon}$, and the remaining pairs obtain the same value. In other words, the effect of collision stays within the immediate neighboring pair.

\subsubsection{Path Graph}
We set $n = 50, g = 0.4, \tau = 0.05, \beta = 4, c = 1, d = 2, \text{ and } \varepsilon = 0.4$. The results indicate that if the collision occurs at the front or the end of the platoon, the $\casr$ will be descending toward the other end of the platoon. If the collision occurs in the middle, the risk obtains a lower value than the previous cases, and it descends toward both ends of the platoon. 

\subsubsection{p-Cycle Graph}
In a $5-$cycle graph and a $10-$cycle graph, we set $n = 50, g = 0.1, \tau = 0.01, \beta = 2, c = 1, d = 2, \text{ and } \varepsilon = 0.1$. The results share some similar properties to the complete graph, i.e., the risk profile also converges to the profile of the complete graph as $p$ increases. The $\casr$ also obtains its peak value at the $p^{th}$ immediate neighbor of the collision pair, this navigates our future research to the
behavior of \eqref{equ:sigma_d} in the eigenspace of a cyclic Laplacian matrix.

\section{Conclusion}
In this work, we develop the value-at-risk measure among cascading systemic events. In a dynamical network of platooning vehicles, the risk of cascading failure (e.g., collision) is quantified using the steady-state distributions, and explicit formulas are obtained for the risk of cascading collisions. 
%In this manner, one can investigate the risk of cascading failures in a networked system regarding the contribution from nodes' and interacting nodes' uncertainties regarding the marginal variance and correlation between the corresponding pairs of vehicles.
Both our theoretical findings and simulations show an evident difference between the risk of single \cite{Somarakis2020b} and cascading systemic failures as the latter one involves the additional information of the pre-occurred failure with its steady-state marginal deviation and correlation between the failed pair of vehicles and the pair of interest.
% Furthermore, we explore risk-based design methods aiming to minimize the risk of cascading failures for certain graphs. For such graphs, the time-delay induced fundamental (lower) limits in the design of graphs with minimum risk of cascading failures are introduced, and a potential generalization of those limits to arbitrary communication graphs. 

A few interesting extensions of the current framework, seen as future work, regard cascading failure events based on multiple existing failures or, to the other extreme, multiple failures being triggered out of a single occurred event. Our future work also focuses on quantifying a fundamental trade-off between network connectivity and the risk of cascading failures, i.e., how by increasing the flow of information within the network, the chances of the domino effect of failures may also increase.

\appendix
\subsubsection{Proof of Theorem \ref{thm:d_steady}}
The affine transformation \eqref{equ:dist} from the normal distributed random variable $\bar{\bm{z}}$ indicates $\bm{\bar{d}}$ follows a multivariate normal distribution, for which the covariance matrix is shown by $\Sigma = D Q \Sigma_z Q^T D^T$. Considering the result from \cite{Somarakis2020b}, we deduce
\begin{equation*}
\sigma_{ij} = \sum_{k=1}^{n} g^2 \big(\tilde{\bm{e}}_{i}^T \bm{q}_k \big) \big(\tilde{\bm{e}}_{j}^T \bm{q}_k \big) \int_{0}^{t} \tilde{\Phi}_{kk} (t-s) ds,
\end{equation*}
then take the limit $t \rightarrow \infty$ to conclude, which presents a generalization of the result in \cite{Somarakis2020b}.\hfill$\square$
    
\subsubsection{Proof of Lemma \ref{lemma:conditional_prob}}
The result follows directly after Theorem \ref{thm:d_steady} and the conditional distribution of a bi-variate normal random variable as in \cite{tong2012multivariate}. \hfill$\square$

\subsubsection{Proof of Theorem \ref{thm:cascade_risk_collision}}
Following Lemma \ref{lemma:conditional_prob}, we can rewrite \eqref{equ:var} as 
\begin{align}
    \inf \Big \{ \; \delta > 0 \; : \; \int_{-\infty}^{ \kappa_{\delta}^{(i,j)}} \exp{\big( -t^2 \big)} \; dt < \sqrt{\pi} \varepsilon\;\Big\}.
\end{align}
The result follows by taking cases on risk value. More specifically, $\mathcal{R}^{C,i,j}_{\varepsilon} = 0$ is equivalent to 
\begin{align*}
    \int_{-\infty}^{\kappa_{0}^{(i,j)}} \exp{\big( -t^2 \big)} \; dt < \sqrt{\pi} \varepsilon ~ \Leftrightarrow  ~ \text{erf}(\kappa_0)< 2 \varepsilon-1
\end{align*}
and the first branch of the formula follows from the invertibility of error function.  Similar argument works for the case $\delta\rightarrow \infty$ by completing the third branch of the risk formula.
   
For the intermediate branch, monotonicity condition on $\kappa_{\delta}^{(i,j)}$ implies there is a unique $\delta^*>0$ such that 
$$\int_{-\infty}^{\kappa_{\delta^*}^{(i,j)}=\iota_\varepsilon} e^{-t^2}\,dt=\sqrt{\pi}\varepsilon. $$ Using the same line of arguments, last equation implies $\kappa_{\delta^*}^{(i,j)}=\iota_\varepsilon$ and the result follows by solving for $\delta^*$. \hfill$\square$

%%%%%%%%%%%%%%%%%%%%%%%%%%%%%%%%%%%%%%%%%%%%%%%%%%%%%%%%%%%%%%%%%%%%%%%%%%%%%%%%%%%%%%%%%%%%%%
%END OF THE MAIN DOCUMENT
%%%%%%%%%%%%%%%%%%%%%%%%%%%%%%%%%%%%%%%%%%%%%%%%%%%%%%%%%%%%%%%%%%%%%%%%%%%%%%%%%%%%%%%%%%%%%%

\printbibliography
\end{document}